\documentclass[a4paper,preprintnumbers,floatfix,superscriptaddress,pra,twocolumn,longbibliography]{revtex4-1}

\usepackage{amsmath, amsthm, amssymb}
\usepackage{graphicx}
\usepackage{dcolumn}
\usepackage{bm}
\usepackage{bbm}
\usepackage{color}
\usepackage[usenames,dvipsnames]{xcolor}
\usepackage{hyperref}
\usepackage{enumerate}
\usepackage{ulem}
\usepackage{textcomp}

\newcommand{\figref}[1]{Fig.~\ref{#1}}

\newcommand{\Tr}{\operatorname{Tr}}

\newcommand{\ba}{\begin{eqnarray}}
\newcommand{\ea}{\end{eqnarray}}

\begin{document}

\title{Practical self-testing quantum random number generator based on an energy bound}

\author{Davide Rusca}
\affiliation{D\'epartment de Physique Appliqu\'ee, Universit\'e de Gen\`eve, 1211 Gen\`eve, Switzerland}
\author{Thomas van Himbeeck}
\affiliation{Laboratoire d'Information Quantique, Universit\'e Libre de Bruxelles, 1050 Bruxelles, Belgium}
\affiliation{Centre for Quantum Information \& Communication, Universit\'e Libre de Bruxelles, Belgium}
\author{Anthony Martin}
\affiliation{D\'epartment de Physique Appliqu\'ee, Universit\'e de Gen\`eve, 1211 Gen\`eve, Switzerland}
\author{Jonatan Bohr Brask}
\affiliation{D\'epartment de Physique Appliqu\'ee, Universit\'e de Gen\`eve, 1211 Gen\`eve, Switzerland}
\affiliation{Department of Physics, Technical University of Denmark, Fysikvej, Kongens Lyngby 2800, Denmark}
\author{Weixu Shi}
\affiliation{College of Electronic Science, National University of Defense Technology, Hunan, Changsha 410073, China}
\affiliation{D\'epartment de Physique Appliqu\'ee, Universit\'e de Gen\`eve, 1211 Gen\`eve, Switzerland}
\author{Stefano Pironio}
\affiliation{Laboratoire d'Information Quantique, Universit\'e Libre de Bruxelles, 1050 Bruxelles, Belgium}
\affiliation{Centre for Quantum Information \& Communication, Universit\'e Libre de Bruxelles, Belgium}
\author{Nicolas Brunner}
\affiliation{D\'epartment de Physique Appliqu\'ee, Universit\'e de Gen\`eve, 1211 Gen\`eve, Switzerland}
\author{Hugo Zbinden}
\affiliation{D\'epartment de Physique Appliqu\'ee, Universit\'e de Gen\`eve, 1211 Gen\`eve, Switzerland}

\begin{abstract}

We present a scheme for a self-testing quantum random number generator. Compared to the fully device-independent model, our scheme requires an extra natural assumption, namely that the mean energy per signal is bounded. The scheme is self-testing, as it allows the user to verify in real-time the correct functioning of the setup, hence guaranteeing the continuous generation of certified random bits. Based on a prepare-and-measure setup, our scheme is practical, and we implement it using only off-the-shelf optical components. The randomness generation rate is 1.25 Mbits/s, comparable to commercial solutions. Overall, we believe that this scheme achieves a promising trade-off between the required assumptions, ease-of-implementation and performance.

\end{abstract}

\maketitle

The device-independent (DI) approach allows for certified quantum random number generation (QRNG) based on minimal assumptions \cite{colbeckPhD,pironio2010,colbeck_private_2011}. 
In particular, no detailed knowledge about the internal working of the quantum devices is needed, and the output can be guaranteed random from the point of view of a hypothetical adversary even in the extreme case where the adversary itself prepared the devices. 
These ideas have generated considerable interest in recent years (see \cite{Acin2016} for a recent review), and first proof-of-principle experiments have been reported \cite{pironio2010,christensen2013,bierhorst_experimentally_2018,liu_device-independent_2018}. 

While conceptually fascinating, the fully DI approach is still at the moment far from being practical since it requires loophole-free Bell tests, which not only are extremely challenging to implement but feature rates in state-of-the-art experiments that are orders of magnitude lower than commercially available QRNGs. 
This has motivated the development of alternative solutions (see e.g. \cite{li2011,vallone2014,lunghi2015,canas2014,Cao2015,Marangon2017,Cao2016,Xu2016,Brask2017a,Gehring2018,Thibault2019}), often referred to as semi-DI (or self-testing), exploring intermediate possibilities between the fully DI setting and the more standard ``device-dependent'' approaches to QRNG, which require a full characterisation of the devices (see e.g. \cite{stefanov2000,jennewein2000,gabriel2010,qi2010,abellan2014,sanguinetti2014}). Such semi-DI solutions are much easier to implement than fully DI protocols, but rely on some extra, even though limited, assumptions on the devices. Though the introduction of these extra assumptions may at first seem a departure from the DI ideal, one should realize that the DI model also requires a certain number of assumptions that are far from being trivially satisfied in practice, such has no information leakage from the devices or that the software used to acquire and process the data functions correctly.
 
In the present work, we report on the implementation of a QRNG, which we believe achieves an excellent trade-off between the required assumptions, performance and ease-of-implementation. The scheme is based on a prepare-and-measure scenario, and thus requires no entanglement or Bell test. The setup features only standard off-the-shelf optical components, and achieves randomness generation rates of the order of MHz, hence comparable to commercial QRNGs. Moreover, the output randomness can be certified based on few natural assumptions. Compared to fully DI model, our scheme requires an additional assumption, namely that the average energy (per signal) of the source is upper bounded. This assumption is arguably quite natural in an optical setup, where mean energy can be directly measured and monitored. Our scheme does not require the assumption of identical and independently distributed runs (i.i.d. hypothesis) and is thus robust to any sort of memory effects. In our experiment we highlight the ``self-testing'' feature of our scheme, which allows the user to verify in real-time the correct functioning of the setup, hence guaranteeing the continuous generation of certified random bits. 



Our scheme fits within the general framework for self-testing (or semi-DI) RNG introduced in \cite{himbeeck2017} and further developed in \cite{himbeeck2018}, which we start by summarising. We consider a prepare-and-measure scenario with a binary input $x$ for the preparation device, and a binary output $b$ for the measurement device. 
For each input, the preparation device sends a quantum state to the measurement device. 
For concreteness, we can take this to describe an optical multimode signal. 
Thus $b$ may depend on $x$, but only via the transmitted quantum states. 
In addition, there may be internal classical noise affecting both the state preparation and the measurements, possibly in a correlated way. 
The observed input-output correlations in a single use of the device can then be written
\begin{equation}
\label{eq.pbx}
p(b|x) = \sum_\lambda p(\lambda) \Tr[ \rho^\lambda_x M^\lambda_b ] ,
\end{equation}
where $\rho^\lambda_x$ are the prepared states, $M^\lambda_b$ are elements of a POVM defining the measurement, and $x,b \in \{0,1\}$, while $\lambda$ is arbitrary and represents the classical noise.

We aim to certify genuinely quantum randomness in the output $b$. 
This means we need to separate any apparent randomness in $b$ due to the classical noise $\lambda$ from that originating from the inherent randomness in the quantum measurements. 
Furthermore, we want to do this with only limited characterisation of the devices. 
In particular, neither the states nor the measurements are known to the user. 
The certification will be based on the input-output correlations $p(b|x)$, together with an assumption about the energy available to encode the quantum states. 
Apart from this assumption, the devices are treated as black boxes (note that writing (\ref{eq.pbx}) we also implicitly assume that the devices do not share prior entanglement).

The energy assumption is formulated as a bound on the average photon number of the quantum light states emitted by the preparation device
\begin{equation}
	\label{eq.energycons}
	\sum_{\lambda} p(\lambda) \Tr[\rho^\lambda_x N] \leq \omega_x \,,
\end{equation}
where $N$ is the multimode photon number operator.

In Ref.~\cite{himbeeck2017}, it was shown that all correlations which can be obtained by mixing deterministic strategies using classical shared randomness must obey
\begin{equation}
\label{eq.classineq}
|p(0|0) - p(1|0) - p(0|1) + p(1|1)| \leq 2(\omega_0 + \omega_1) .
\end{equation}
We now introduce a physical setup in which this bound can be violated. 
Such a violation implies that the devices' behaviour cannot be explained deterministically and thus that genuine quantum randomness is produced at the output of the measurement device.

\begin{figure}[t]
  \centering
  \includegraphics[width=\columnwidth]{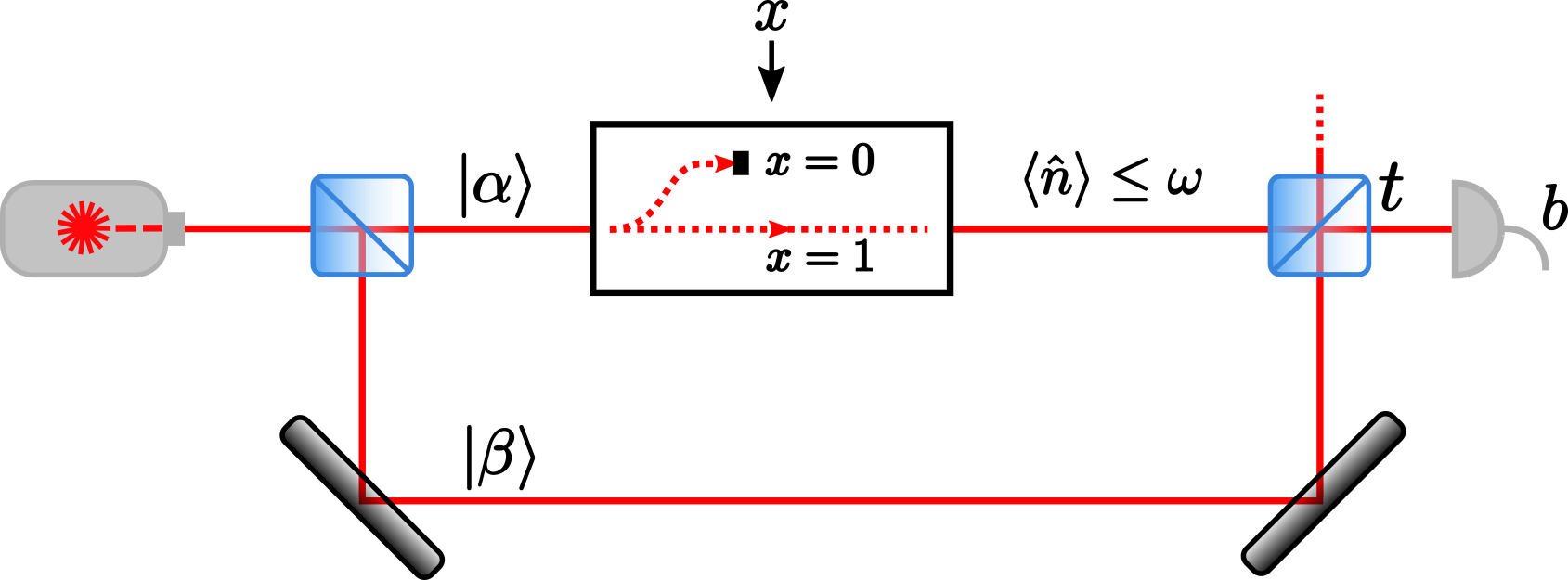}
  \caption{Self-testing quantum random number generation based on bounded energy. 
A signal coherent state is modulated depending on the input $x$. 
The signal is blocked for $x=0$ and transmitted for $x=1$ and the average energy after the modulation is bounded by $\omega$. 
The signal is mixed with a local oscillator on a beam splitter with transmitivity $t^2$ and measured by a single-photon detector.}
  \label{fig.schemetheory}
\end{figure}

The setup is show in \figref{fig.schemetheory}. 
Depending on the input $x$, the amplitude of a signal coherent state (produced by a laser) is modulated such that the output amplitude for $x=0$ is 0, while for $x=1$ it is $\alpha$. 
The transmitted state is then measured by interfering it with a local oscillator with amplitude $\beta$ on a beam splitter, followed by single-photon threshold detection in one output port. 
The other output port is simply ignored. 
The beam splitter has transmitivity $t^2$ and reflectivity $r^2 = 1-t^2$. 
In the event the detector does not click, we assign the output $b=0$, while $b=1$ corresponds to a click. 
The probabilities for the detector not to click can be computed explicitly
\begin{align}
p(0|0) & = e^{-\eta|r\beta|^2} , \\
p(0|1) & = e^{-\eta|t\alpha+r\beta|^2} ,
\end{align}
where $\eta \in [0,1]$ represents a combined transmission and detection efficiency accounting for losses and inefficient detectors \footnote{Since the local oscillator amplitude $\beta$ is completely free, if the losses are different in each arm, we can just absorb a factor into $\beta$ to make up for this difference.}. 
The remaining probabilities are determined by normalisation.

The mean photon numbers  at the output of the preparation device are
\begin{equation}
\langle N \rangle = \begin{cases}
0 & \text{for} \,\, x=0 \\
|\alpha|^2 & \text{for} \,\, x=1\,,
\end{cases}
\end{equation}
and we can then take the energy bounds on the transmitted states \eqref{eq.energycons} to be equal to the these mean photon numbers. 
Note that the local oscillator carries no information about $x$ and is not considered to be part of the prepared state. 
In particular, no assumption is made on the amount of energy in the local oscillator arm.

In our case the inequality \eqref{eq.classineq} then becomes
\begin{equation}
\label{eq.setupineq}
|e^{-\eta|t\alpha+r\beta|^2} - e^{-\eta|r\beta|^2}| \leq |\alpha|^2 ,
\end{equation}
It is easy to see that for any value of $\eta>0$, there exist choices of $\alpha$ and $\beta$ for which \eqref{eq.setupineq} is violated. 
For instance, take $\alpha=\eta t/2$, $\beta=1/r$. 
Then one can verify analytically that the inequality is violated for small $\eta$ by expanding the left-hand side in $\eta$, and one can check numerically that it is also the case for all larger $\eta$. 
Thus, for any non-zero efficiency, our scheme admits $\alpha$ and $\beta$ for which the output $b$ is not deterministic.


Given data that violates \eqref{eq.classineq}, we need to quantify the randomness generated by the measurement device. 
For a device behaviour which is independent and identically distributed (i.i.d.) in each experimental round, the optimal asymptotic rate of randomness generation, relative to an observer with knowledge of the input $x$ as well as the internal variable $\lambda$, is given by the Shannon entropy $H(B|X,\Lambda)=-\sum_{b,x,\lambda} p(\lambda) p(x)p(b|x)\log_2 p(b|x,\lambda)$ (this follows from the asymptotic equipartition theorem \cite{cover_elements_2006}). 
For any given values of the probabilities $p = \{p(x,b)\}_{b,x}$ and the energy bounds $\omega = \{\omega_x\}_x$ (or linear functions of them), it was shown in \cite{himbeeck2018}, how using semidefinite programming one can put a lower-bound on $H(B|X,\Lambda)$ that is valid for arbitrary decompositions \eqref{eq.pbx}, \eqref{eq.energycons} that use hidden classical noise $p(\lambda)$.

\begin{figure}
	\includegraphics[scale=0.8]{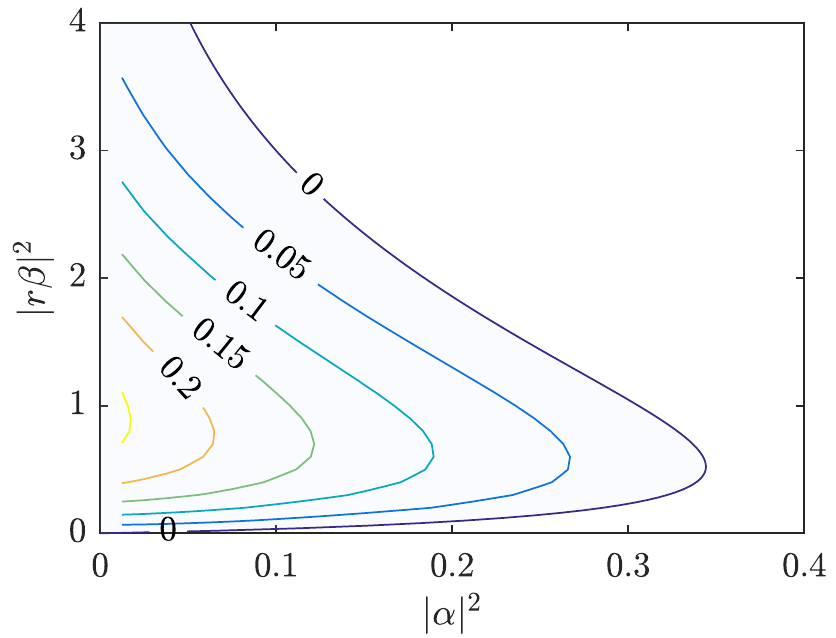}
	\caption{Asymptotic rate of randomness generation (in bits) as a function of the experimental parameters. 
The rate is the worst-case entropy $H(B|X\Lambda)$, given the correlations $p$ and the average energy constraint $\bar \omega$. 
These were obtained assuming an identical phase $\arg(\alpha) = \arg(\beta)$, a biased input distribution $p_x(1) = 0.25$, detection efficiency $\eta = 50\%$ and transmission $t^2 = 99\%$. 
Only the $\alpha$ and $\beta$ in the shaded region satisfy the inequality \eqref{eq.setupineq}. 
	\label{fig.entropy}}
\end{figure}

As an illustration, we show in \figref{fig.entropy}, the entropy for several values of $\alpha$ and $\beta$, using $\eta=50\%$, $t^2 = 99\%$, a biased input distribution $p(1) = 25\%$, and assuming the bound $\sum_x p(x) \omega_x \leq \bar\omega=p(1)|\alpha|^2$ on the average value of the energies $\omega_x$.

The SDP introduced in \cite{himbeeck2018} not only returns the lower-bound on $H(B|X,\Lambda)$, but also provides a self-test or witness certifying this amount of randomness. 
This self-test consists of a linear function $\gamma[p]-\zeta[\omega]$ in $p$ and $\omega$ with the property that $H(B|X,\Lambda)\geq \gamma[p]-\zeta[\omega]$. 
Once such witness is known it thus suffices to evaluate it on $p$ and $\omega$ to obtain a lower-bound on $H(B|X,\Lambda)$.

The existence of such a witnesses, which can be computed for any given $p$ and $\omega$ immediately suggest a semi-DI RNG protocol: (1) fix a certain witness (tailored to the expected behavior of the devices); (2) run the devices $n$ times and record the inputs $\bm{X} = (X_1,X_2 \cdots)$ and outputs $\bm{B}= (B_1,B_2\cdots )$; (3) compute the value $\gamma[f]$, where $f(x,b)$ are the frequencies of occurrence of $(X_i,B_i)=(x,b)$, and check if $\gamma[f]- \zeta[\omega]\geq h$ is above some threshold $h$. 
The passing of this test, denoted by $\mathrm{Pass}$, establishes that the device is working properly.

In the implementation of such a protocol, it is not necessary the case that the device behaves in a i.i.d way and its general behavior is now described by an unknown $n$-round joint distribution $p_{[\bm{B},\bm{X},\bm{\Lambda}]}$. 
Still the observed value of the linear witness, computed from the observed frequencies, certifies a certain amount of randomness in the output string $\bm{B}$. 
Specifically, in \cite{himbeeck2018}, it was shown that the distribution $p_{[\bm{B},\bm{X},\bm{\Lambda}|Pass]}$ conditioned on the passing of the test contains an amount of randomness given by
\begin{equation}
	\label{eq.finite_rate}
	H^{\epsilon'}_{\min}(\bm{B}|\bm{X},\bm \Lambda)\geq n \left(h - c \sqrt{\tfrac{\log(\epsilon/2)}{n}} - d \tfrac{\log(\epsilon/2)}{n}\right)\,.
\end{equation}
Here $H_{\min}^{\epsilon'}(\bm{B}|\bm{X},\bm{\Lambda})$ is the worst-case conditional smooth min-entropy, defined as the largest $k$ such that $\Pr( -\log_2 p(\bm B|\bm X \bm \Lambda \mathrm{Pass}) \geq k) \geq 1-\epsilon'$. 
It roughly corresponds to the largest number of bits that can be extracted from the output string $\bm{B}$ using a strong extractor, such that the resulting distribution deviates from an ideal distribution (i.e., uniform and independent of $X\Lambda$) at most with probability $\epsilon'$. 
The security parameter $\epsilon$ in the right-hand side of (\ref{eq.finite_rate}), which should be chosen small ($\epsilon = 10^{-10}$ in the following), is related to the smoothing parameter though $\epsilon' = \epsilon/\Pr(\mathrm{Pass})$. 
This ensures that the protocol is $\epsilon$-sound, because the probability of both passing the test and deviating from an ideal distribution $\Pr(\mathrm{Pass})\times\epsilon' = \epsilon$ is guaranteed to be small. 
The subleading error terms in \eqref{eq.finite_rate}, given by the constants $c,d>0$, depend on the choice of witness $\gamma[.]$ and $\zeta[.]$ and are given in \cite{himbeeck2018}.


\begin{figure}[t]
  \centering
  \includegraphics[scale=0.5]{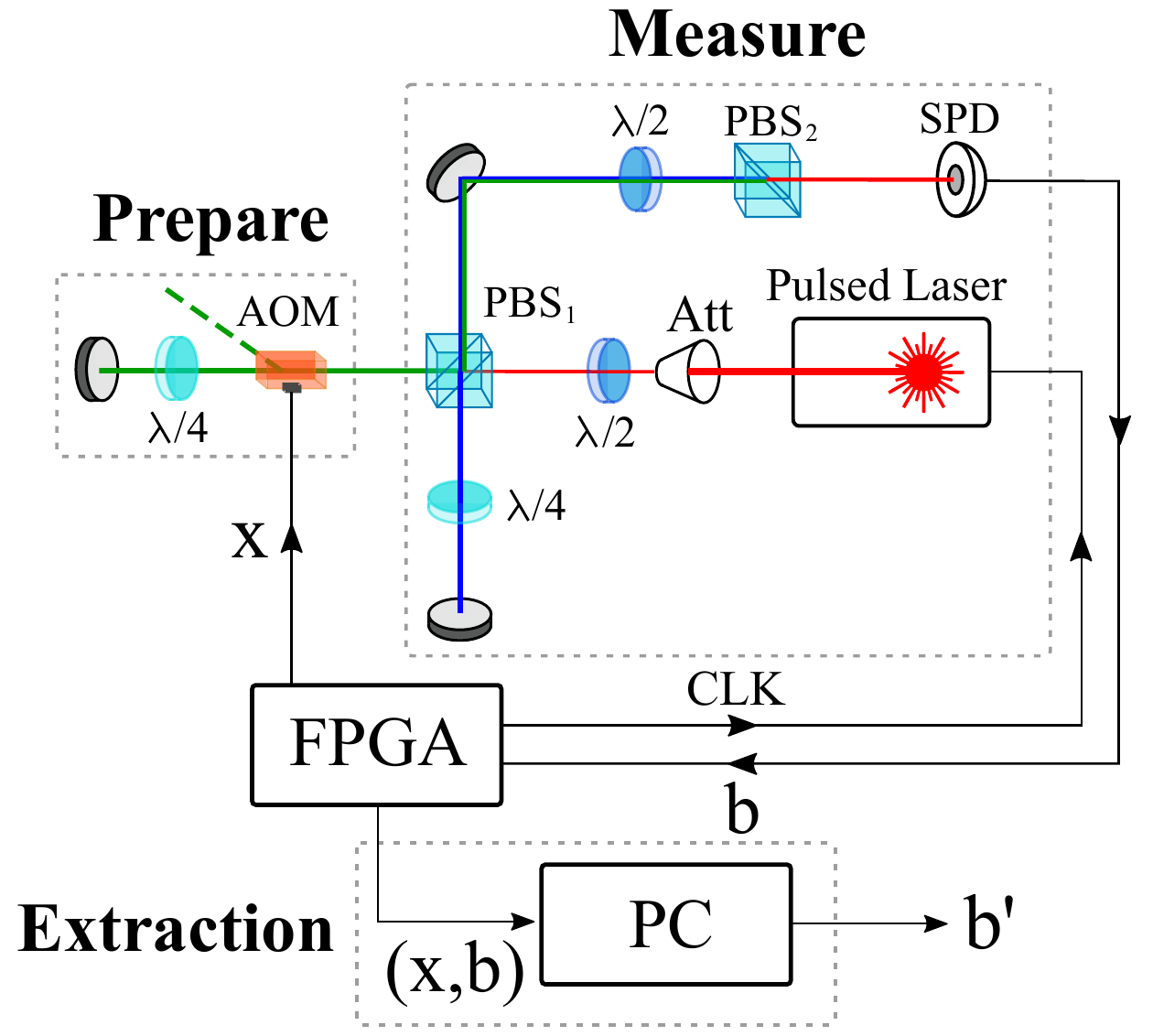}
  \caption{Experimental setup. 
A coherent stage is generated by a pulsed diode laser and send to a Michelson interferometer. 
On one arm (green path) is used to prepare the signal and the other one (blue path) to prepare the local oscillator.}
  \label{fig.setup}
\end{figure}

In order to implement the scheme presented in \figref{fig.schemetheory} we used the experimental set-up drawn in \figref{fig.setup}.
A coherent state is generated by a pulsed diode laser at 655\,nm trigged by a field-programmable gate array (FPGA) at 12.5\,MHz. 
A set of half-wave plate (HWP) and polarizing beam-splitter (PBS) is used to prepare the local oscillator ($\beta$) and the signal ($\alpha$) and to tune the amplitude ratio between them. 
To maximize the transmission to the output port of the interferometer, one quarter-wave plate (QWP) is inserted in each arm to rotate the polarization of the incoming light. 
 For each pulse, a pseudo-random binary input $x$ is generate by the FPGA and sent to the acousto-optic modulator (AOM). 
When $x=1$ the AOM deflects the light which introduce an additional losses of more than 23~dB.

At the output of $\text{PBS}_1$, the signal and local oscillator are in the same spatial and temporal mode but have orthogonal polarizations. 
To make them interfere they pass through a HWP and $\text{PBS}_2$. 
The HWP is oriented to achieve a transmission of $t^2 = 99\%$ for the signal. 
Then, the light is detected by a single photon detector (SPD) (PerkinElmer SPCM-AQR) with an efficiency of 77\% for a dark-count rate around 300Hz and the digital signal is recorded by the FPGA. 
Each second the FPGA sends the data to a personal computer for storage.

In order to bound the average energy on the signal, the power was measured at the output of $\text{PBS}_1$ while obstructing the local oscillator arm. 
This was done with a linear power meter (Thorlabs PM100D with S122C sensor) with an absolute uncertainty of $\pm 5\%$. 
The average energy bound was then calculated by considering the attenuation chosen before the interferometer and by dividing the resulting power by the repetition rate. 
The power on the signal and local oscillator was adjusted in order to maximise the amount of entropy generated based on results presented in \figref{fig.entropy}. 
These measurements were done twice, before and after the experiment, in order to ensure the stability of the power. 
Alternatively, the power could be monitored in real time using a beam splitter between the two PBSs, in a similar manner as in Ref.~\cite{Brask2017a}.

\begin{figure}[t]
  \centering
  \includegraphics[width=\linewidth]{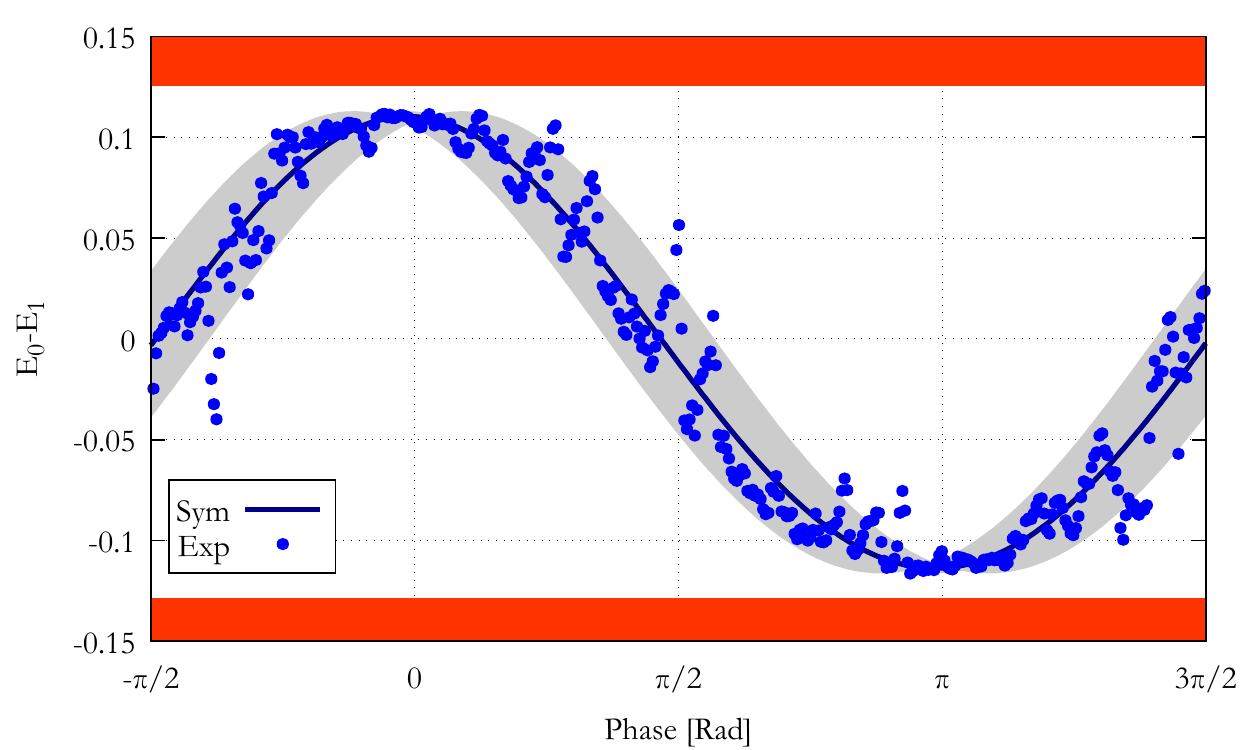}
  \caption{Correlations function with respect to the phase difference between the two arms of the interferometer. 
The solid line corresponds to the simulated behaviour of the device without phase fluctuation. 
The grey area correspond to a relative phase incertitude region of $\pm 10\%$. 
The red area corresponds to the region of correlations unachievable once the bound on the energy $\omega$ is chosen to be $0.0025$.}
  \label{fig.E}
\end{figure}
After the energies of the signal and local oscillator are set, we launch a measurement to estimate the correlation between the modulation of the signal and the detections.
The measurement is carried out by changing the relative phase between $\alpha$ and $\beta$. 
This is done by moving the mirror in the local oscillator arm with a piezoelectric translator.
The relevant marginals are then estimated as $\overline{p}(b|x) = n_{b,x}/n_x$, where $n_{b,x}$ and $n_x $ correspond to the number of events for an output $b$ given an input $x$ and the total number of state $x$ send, respectively.
Then, we calculated the correlation function $E~=~p(b = x) - p(b \neq x)$ shown in \figref{fig.E} as function of the phase difference between $\alpha$ and $\beta$ with an $\omega_{\rm estimate} = 0.0022(1)$. 
The energy of the local oscillator $\beta$ is set to $99$ mean photon per pulse which corresponds to  $(1-t^2)|\beta|^2 = 0.99$ with $1-t^2 = 0.01$.
The solid line in the figure corresponds to the theoretical prediction for a perfect system with a signal state with mean photon number of  $\omega = 0.0025$ and global detection efficiency of $55\%$. 
As it is shown in the figure the experimental points do not achieve the same optimal points with respect to the theoretical expectation. 
This difference is due to the fact that $\omega_{\rm estimate} <  \omega$ in order to avoid the eventuality of violating the general assumption. 
The fluctuation in \figref{fig.E} are due to fluctuations of the relative phase between the signal and local oscillator owing to the instability of the interferometer. 

To perform a QRNG protocol, we considered 35 blocks of $n = 10^8$ rounds ($8~s$ measurement) in which the interferometer was around a constructive interference. 
A witness was determined and optimized for this regime by using one trial block and, assuming an average energy assumption $\bar \omega = 0.0025$, a threshold was chosen at the value $h = 0.117$, corresponding to a rate \eqref{eq.finite_rate} of $0.1$ bits/round. 
The test was satisfied in all blocks, giving an average output rate of certified quantum randomness of 1.25~MHz. 
A possible improvement on our experiment could be achieved by actively stabilizing the interferometer, procedure that will increase both the extractable entropy and the passing probability.

In conclusion, we developed a simple scheme for a self-testing QRNG based on an energy bound. This scheme represents in our opinion an excellent compromise between the required assumptions, experimental complexity and performance. From the point of view of security, we believe that weakening the required assumptions without moving to the full DI scenario will be difficult. From the point of view of implementation and performance, progress can still be achieved, for instance by using integrated optics.

\emph{Acknowledgements.---} We acknowledge support from the Swiss National Science Foundation (Starting grant DIAQ, Bridge project ``Self-testing QRNG'' and NCCR-QSIT), the EU’s H2020 program under the Marie Sklodowska Curie project QCALL (GA 675662) and the EU Quantum Flagship project QRANGE.

\bibliography{maxavg_qrng}

\appendix

\end{document}